\documentclass[dvipsnames]{aastex63}
\usepackage[version=4]{mhchem}
\usepackage{amsmath}

\received{\today}
\revised{\today}
\accepted{\today}
\submitjournal{ApJL}

\shorttitle{Water shielding and Organics}
\shortauthors{Duval et al}

\begin{document}
\title{Water shielding in the terrestrial planet-forming zone: Implication for inner disk organics}
\date{\today}

\correspondingauthor{Sara Duval} 
\email{saraduval018@gmail.com}

\author[0000-0003-0014-0508]{Sara E. Duval}
\affiliation{University of Michigan, LSA astronomy \\
1085 S. University Avenue \\
Ann Arbor, MI 48109, USA}

\author[0000-0003-4001-3589]{Arthur D. Bosman}
\affiliation{University of Michigan, LSA astronomy \\
1085 S. University Avenue \\
Ann Arbor, MI 48109, USA}

\author[0000-0003-4179-6394]{Edwin A. Bergin}
\affiliation{University of Michigan, LSA astronomy \\
1085 S. University Avenue \\
Ann Arbor, MI 48109, USA}

\begin{abstract}

The chemical composition of the inner region of protoplanetary disks can trace the composition of planetary building material. The exact elemental composition of the inner disk has not yet been measured and tensions between models and observations still exist.
Recent advancements have shown UV-shielding to be able to increase emission of organics. Here, we expand on these models and investigate how UV-shielding may impact chemical composition in the inner 5 au.
In this work, we use the model from \citet[][]{Bosman2022water} and expand it with a larger chemical network. We focus on the chemical abundances in the upper disk atmosphere where the effects of water UV-shielding are most prominent and molecular lines originate. 
We find rich carbon and nitrogen chemistry with enhanced abundances of \ce{C2H2}, \ce{CH4}, HCN, \ce{CH3CN}, and \ce{NH3} by $>$ 3 orders of magnitude. This is caused by the self-shielding of \ce{H2O}, which locks oxygen in water. This subsequently results in a suppression of oxygen-containing species like CO and \ce{CO2}. The increase in \ce{C2H2} seen in the model with the inclusion of water UV-shielding allows us to explain the observed \ce{C2H2} abundance without resorting to elevated C/O ratios as water UV-shielding induced an {\em effectively} oxygen-poor environment in oxygen-rich gas.
Thus, water UV-shielding is important for reproducing the observed abundances of hydrocarbons and nitriles. From our model result, species like \ce{CH4}, \ce{NH3}, and NO are expected to be observable with the James Webb Space Telescope (JWST). 

\end{abstract}





\section{Introduction} \label{sec:intro}

\begin{indent}
Within protoplanetary disks, the inner 2 to 3 au is a critical location in which much of the process of planet formation is believed to occur, referred to as the planet-forming zone \citep[e.g.][]{Pierrehumbert2011, Morbidelli2012, Raymond2014, Morbidelli2016}. Most directly, this region corresponds to radii at which terrestrial planets are formed \citep{Mulders2015, Madhusudhan2021}. Observations have shown that many stars are expected to have a planet within 1 au \citep[][]{Johnson2010, Mulders2018}. The inner 1 au is inside or contains the \ce{H2O} iceline and water plays a large role in the evolution of life \citep[e.g.][]{Brown2013, Cockell2016, Lingam2019}. Furthermore, inside the water iceline the elemental C/O ratio is predicted to be $\sim$ stellar, which becomes inherited by giant planets \citep{Oberg2011, Ida2019, Oberg2021}.  Thus, the location of formation impacts chemical composition \citep[e.g.][]{Lahuis2006, Oberg2011, Pontoppidan2011, Walsh2012, Madhusudhan2019}.

The inner disk is dust-rich leading to high optical depths. This makes it difficult to determine chemical content of the disk midplane. However, the radiation from the star, which is in close proximity, leads to a heated disk surface (i.e. T$_{gas}$ $>$ T$_{dust}$), which produces a rich spectrum of emission from volatile molecules, particularly at infrared wavelengths \citep[e.g.][]{Carr2008, Brown2013}. Water vibrational lines at 3 $\mu m$, HCN vibrational at 3.3 $\mu m$, and CO vibrational at 4.7 $\mu m$ have been observed with both Keck Near Infrared Spectrograph (NIRSPEC) and VLT-CRIRES \citep[e.g.][]{SalykCO2011, Brown2013, Mandell2013, Carr2018}.  Further lines of \ce{H2O}, OH, HCN, \ce{C2H2}, and \ce{CO2} have been detected by the Spitzer Space Telescope Infrared Spectrograph (IRS) which ranges from 10-37 $\mu$m \citep[e.g.][]{Carr2008,Salyk2008,Salyk2011,Pontoppidan2010}. All these lines are thought to originate in the inner 2 to 3 au,  which allows us to trace the gas composition  
These observations show that a diverse chemistry is present in the inner planet-forming zone.


\citet[][]{Najita2013} and \citet[][]{Najita2018} investigate the ratio of HCN/\ce{H2O} line flux and find it is related to disk dust mass. They argue that this relation is due to the formation of planetesimals which decouple from the dust and lock up water in distant regions. This effectively increases the C/O ratio in the inner disk directly impacting the HCN/\ce{H2O} line flux. \citet[][]{Banzatti2020} explores the same dataset and find that the strongest relation is an anti-correlation between L$_{\ce{H2O}}$ and R$_{dust}$. Instead of an elevated C/O, they propose that the inner disk is fed by drifting pebbles, where large disks are a sign of little drift and small disks are a sign of substantial drift. Since these pebbles are water-ice rich, a high drift rate will enhance the inner disk oxygen content when the water ice sublimates. To distinguish these scenarios we need to understand the C/O ratio of inner disk gas.

One way to determine the inner disk chemical content and C/O ratio is to use detailed thermo-chemical models. Models such as Dust and LInes (DALI) \citep{Bruderer2012, Bruderer2013}, RAC2D \citep{Du2014}, and Protoplanetary Disk Model (ProDiMo) \citep{Woitke2009}, solve for both the gas thermal physics and the chemical equilibrium, given stellar parameters, the dust properties and mass distribution, and an overall gas-to-dust mass ratio. 
Based on these models, \citet[][]{Woitke2018} and \citet[][]{Anderson2021} find that altering the C/O ratio changes the predicted emission of molecules arising from inner disk gas. In their models, they have matched emission from multiple organics, but underproduced \ce{C2H2} unless an elevated C/O ratio is invoked.

\citet{Bethell2009} showed that strong formation rates of water vapor in hot ($>$400~K) surface gas can compete with ultraviolet (UV) photodestruction, allowing water to self-shield.  Since water has a broad UV photoabsorption cross-section \citep[][]{Yoshino96}, this can also shield other molecules downstream, a process called water UV-shielding. \citet{Bosman2022water} (hereafter, Paper \MakeUppercase{\romannumeral 1}), \citet[][]{Calahan2022H218O} (Paper \MakeUppercase{\romannumeral 2}), and \citet{Bosman2022CO2} (Paper \MakeUppercase{\romannumeral 3}) have shown that water UV-shielding lowers the UV flux deeper into the disk and is important in understanding H$_2$O, H$_2^{18}$O, and \ce{CO2} emission. 

Additional species beyond \ce{H2O} such as: \ce{H2}, CO, \ce{CI} , \ce{HI} and \ce{N2} are also abundant in the surface layers to potentially impact the UV field. \ce{H2}, CO, atomic carbon, and \ce{N2} all absorb wavelengths less than 110 nm, while most other species dissociate at wavelengths $>$ 110 nm and most of the UV photons are also in this wavelength range \citep[e.g.][]{Herczeg2004, Heays2017}. Thus, while the UV attenuation of these species greatly impacts each other, they do not greatly impact the dissociation of other species, in contrast to \ce{H2O}. Shielding by atomic H, specifically scattering of Ly-$\alpha$ does not seem to have a big impact of the chemistry of the inner disk (Paper \MakeUppercase{\romannumeral 2}). 

This naturally raises the question as to whether water UV-shielding affects the rest of the chemistry in the inner disk. 
This is what we investigate in this paper: the impact of water UV-shielding on chemistry in the inner disk, with the goal of reconciling current models with observations and making predictions for observations with the James Webb Space Telescope.

\end{indent}

\section{Methods} \label{sec:Meth}
\begin{table}[]
    \centering
        \caption{Disk model parameters}
    \begin{tabular}{l|c}
    \hline \hline
Parameter &  Value \\
\hline
Stellar Luminosity &1 $L_\odot$ \\
Stellar Spectrum & AS 209 $^{a}$\\
Stellar Mass & $ 1.0 M_\odot$ \\
X-ray luminosity & $10^{30}$ erg s$^{-1}$\\
Cosmic ray ionization rate& $10^{-17}$ s$^{-1}$\\
Sublimation radius & 0.08 AU \\
Critical radius & 46 AU \\
Disk outer radius & 100.0 AU \\
Gas surf. dens. at $R_c$ & 21.32 g cm$^{-2}$ \\
Surf. dens. powerlaw slope & 0.9 \\
Disk opening angle & 0.08 \\
Disk flaring angle & 0.11 \\
Large dust fraction& 0.999 \\
Large dust settling& 0.2 \\
\hline
        Element & Abundance w.r.t. H \\
    \hline
        H &  1.0 \\
        He & $7.59 \times 10^{-2}$ \\
        C & $1.35 \times 10^{-4}$ \\
        N & $2.14 \times 10^{-5}$ \\
        O& $2.88 \times 10^{-4}$ \\
        Mg& $4.17 \times 10^{-9}$ \\
        Si& $7.94 \times 10^{-8}$ \\
        S& $1.91 \times 10^{-8}$ \\
        Fe& $4.27 \times 10^{-9}$ \\
    \hline
    \end{tabular}

    \label{tab:Chem_elem_abu}
\end{table}
\subsection{Model setup}

We use the DALI models from Paper \MakeUppercase{\romannumeral 1}. These models include modification from standard DALI \citep{Bruderer2012, Bruderer2013} to better represent the inner disk regions. These include more efficient \ce{H2} formation at high temperature, more efficient heating following photo-dissociation \citep[following][]{Glassgold2015} and water UV-shielding \citep[][Paper \MakeUppercase{\romannumeral 1}]{Bethell2009}. The models have an AS 209 like input spectrum with most of the UV in Ly-$\alpha$ taken from \citet{Zhang2021MAPS}. Model setup details are in Paper \MakeUppercase{\romannumeral 1} and model parameters are reiterated in Table~\ref{tab:Chem_elem_abu}. For this Letter we will only focus on one of the four structures discussed in Paper \MakeUppercase{\romannumeral 1}, the flat ($h_c$ = 0.08) model with a gas-to-dust ratio of $10^5$ as this model is best able to reproduce both the water and \ce{CO2} emission (Paper \MakeUppercase{\romannumeral 3}). The elemental abundances assumed in the chemistry are found in Table~\ref{tab:Chem_elem_abu} and are based on \citet{Jonkheid2006}, with reduced Mg, Si, S and Fe. Finally, as the chemical time-scale in the region of interest are short, we solve for statistical equilibrium  \citep{Anderson2021}. 

The DALI models use a simplified chemical network, which is sufficient for species such as \ce{CO}, \ce{H2O} and \ce{CO2}. However, it does not include the reactions for realistic abundances of the organics, including \ce{C2H2} and \ce{HCN}, which have been commonly observed with \textit{Spitzer}-IRS \citep{Carr2008, Pontoppidan2010}. To correct this, we use an expanded chemical network. This network is based on the network from \citet{Walsh2015} and includes modifications as denoted in \citet{Bosman2021aMAPS}. Furthermore, we made sure that the adaptations made to the simplified network, such as the \ce{H2} formation reactions, 3-body reaction, and the collisional dissociation reactions are correctly incorporated into the bigger chemical network. 
This chemical network is then used to calculate the chemical composition with the gas-temperature and UV field from the standard DALI model. We adopt the \ce{H2O} abundances from the full chemical calculation to calculate the shielding of UV photons for the chemistry, but we do not update the gas temperature. More details on the model setup can be seen in Appendix~\ref{app:appendix}. Figure~\ref{fig:2D_9} contains information about the disk structure for our UV-shielding model. In Fig.~\ref{fig:2D}, we show that the water abundances between the simplified and full network are very similar in the surface layers where water UV shielding is most important.

\subsection{Determination of emitting layer}

In Paper \MakeUppercase{\romannumeral 1}, hot ($>$300--400~K) water vapor is found in high abundance in a thin surface layer that is radially confined within 1 au. In this region, the water vapor column exceeds 10$^{20}$ cm$^{-2}$ and with a gas-to-dust ratio of 10$^{5}$, dust UV absorption is negligible compared to water UV shielding. Thus, including water shielding alters the UV transfer in the surface of the disk which should lead to large differences in our models.
In the disk surface, UV photons heat the gas, creating T$_{gas}$ $>$ T$_{dust}$, a condition necessary for the emission of molecular lines. Inclusion of UV shielding thus is expected to lower the gas temperature and decrease emission deeper in the disk. This provides a boundary from below which we do not expect emission. In Papers \MakeUppercase{\romannumeral 1} - \MakeUppercase{\romannumeral 3}, we found that the emission comes from gas of T$_{gas}$ $\gtrapprox $ 300-400 K and a hydrogen nuclei column of $\	\lessapprox $ 10$^{24}$ cm$^{-2}$. This paper does not explore the radiation transfer, so we consider the top 10$^{24}$ cm$^{-2}$ as our ``infrared (IR) emitting layer'', corresponding to a z/r $\gtrsim$ 0.15, and focus our analysis on the chemical composition of this layer.

\section{Results}
\label{sec:Res}
\begin{figure*}
  \centering
    \includegraphics[width = \hsize, trim=0in 0.7in 1in 1in, clip]{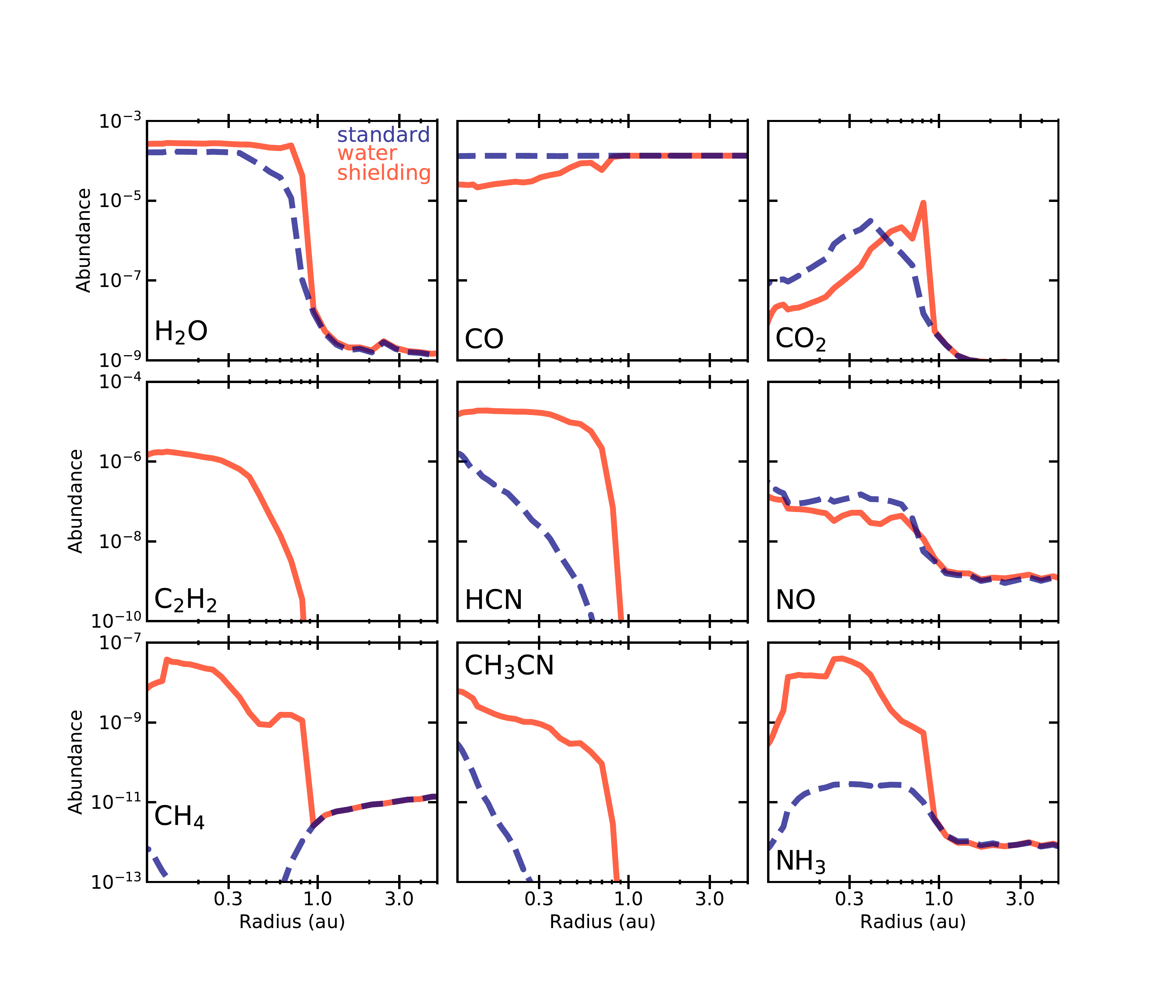}
    \caption{The molecular abundance of various species over the inner 5 au, determined from the IR emitting layer, for the standard and water shielding models. The locations of the water and \ce{CO2} icelines are at 0.27 au and 0.94 au, respectively. The drops in abundance near 0.9 au are a result of drops in the gas temperature in the surface layer (See.~\ref{fig:2D_9}).}
    \label{fig:mol_abuIRlayer}
\end{figure*}
    
     Figure~\ref{fig:mol_abuIRlayer} shows the abundance in the emitting layer for two models, with and without water UV shielding (water shielding and standard, respectively). For \ce{H2O}, the predicted abundance remains relatively similar at radii within the \ce{CO2} iceline, where the model with UV-shielding has more abundant water. In the case of CO, there is a clear depletion in abundance for the model with water UV-shielding. In the disk surface, CO can be destroyed by two pathways. In the highest reaches where CO is not fully shielded, it can be directly dissociated by UV radiation or through reactions with He$^{+}$, which requires ionization by X-rays. Both pathways create free oxygen which is stolen to make \ce{H2O} before CO can reform. In summary, the important reaction pathway is:
     \begin{equation}
    \begin{split}
        \ce{CO + h$\nu$ -> C + O} \\
        \ce{He$^{+}$ + CO -> C$^{+}$ + O + He} \\
        \ce{O + \ce{H2} -> OH + H} \\
        \ce{OH + \ce{H2} -> \ce{H2O} + H}. 
    \end{split}
    \end{equation}
     which leaves carbon without any oxygen to reform CO.
     This excess carbon is sequestered in large (hydro)carbon-chain species, which will be discussed in Sec.~\ref{ssc:carbdisc}.
     
     The behavior of \ce{CO2} varies with radius. In the inner 1 au, the progressively lower temperature at increasing radius increases the \ce{CO2} formation rate relative to the water formation rate \citep[e.g.][]{Bosman2018}. Outside of the \ce{CO2} midplane iceline, the temperatures of the surface layers become too low for efficient formation of OH, the main precursor of \ce{CO2}, causing a strong \ce{CO2} abundance drop.  \ce{H2O}, \ce{C2H2}, HCN, NO, \ce{CH4}, \ce{CH3CN}, and \ce{NH3} all experience a similar temperature-driven drop in abundance around 0.9 au.
    
    We have selected a number of species that have a many order of magnitude difference in abundance, between the standard model and including water UV-shielding, out to the \ce{CO2} iceline: \ce{C2H2}, HCN, \ce{CH4}, \ce{CH3CN}, and \ce{NH3}. For these species, water UV shielding has a significant effect, enhancing these abundances by $>$ 3 orders of magnitude, in disks with high gas-to-dust mass ratios, where there is significant dust settling and growth. We expect that the species with an enhanced abundance due to water UV-shielding will be able to be observed more prominently at heights further up in the disk atmosphere. Figure~\ref{fig:abundantspec} shows the emitting columns for these species at 0.3 au and 0.6 au along with the other abundant species that are present in our UV-shielded model. It can be seen that the impact of UV-shielding on molecular abundance yields significant results, impacting abundance at radii less than 0.9 au.

\begin{figure*}
\begin{minipage}[c]{\hsize}
    \vspace*{-0.5 cm}
    \includegraphics[width = \hsize, trim=1in 0in 1in 0in, clip]{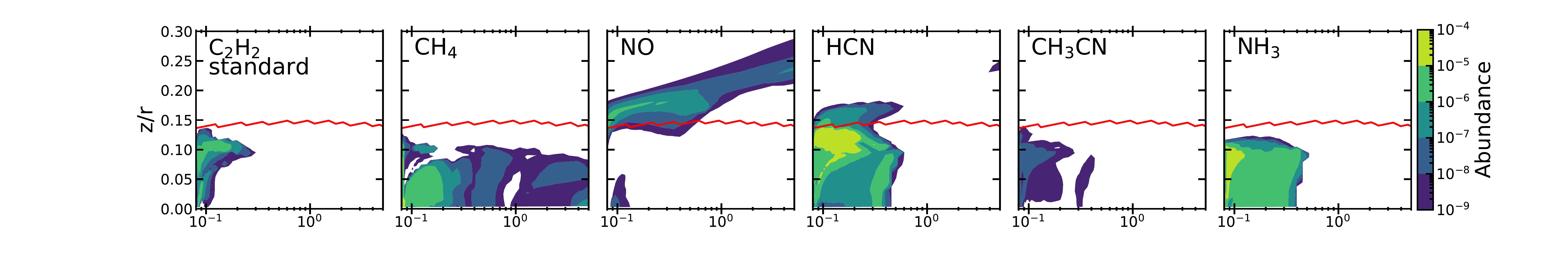}
\end{minipage}
\begin{minipage}[c]{\hsize}
    \vspace*{-0.5 cm}
    \includegraphics[width = \hsize, trim=1in 0in 1in 0in,
    clip]{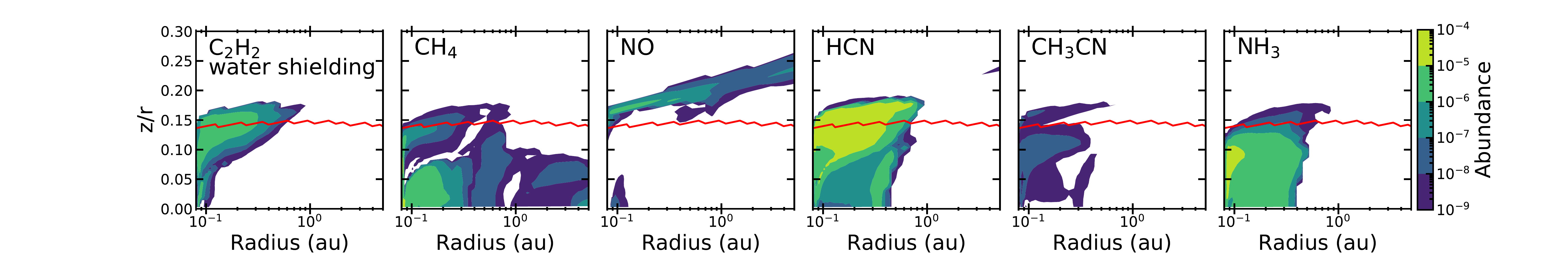}
\end{minipage}
    \caption{The 2D abundance distributions of the 6 species for the standard and water shielding models in the inner 5 au. The red line signifies the location of our estimate of the IR emitting layer above z/r $\sim$ 0.15. It is clear that with the inclusion of water UV-shielding, these species reach higher abundances at locations higher up in the disk atmosphere.}
    \label{fig:2Dspecies}
\end{figure*}

Figure~\ref{fig:2Dspecies} shows the 2D abundance distributions for the standard model and the water shielding model over the inner 5 au for \ce{C2H2}, \ce{CH4}, NO, HCN, \ce{CH3CN}, and \ce{NH3}. Looking in the IR emitting layer, the area above the red line at z/r $\sim$ 0.15, it is clear that this region is most affected by the inclusion of water UV-shielding. We see that \ce{C2H2}, \ce{CH4}, \ce{CH3CN}, and \ce{NH3} are found in the IR emitting layer only if water UV-shielding is included, while NO becomes depleted relative to the standard model, as was found in the average abundances in Fig.~\ref{fig:mol_abuIRlayer}. At deeper layers, below the IR emitting layer, practically identical results are seen for both models, reflecting that the differences in abundance are limited to the surface layers. Thus, Fig.~\ref{fig:mol_abuIRlayer} includes the entire vertical column of gas that is impacted by UV-shielding.

\begin{figure*}
  \centering
    \includegraphics[width = \hsize, trim=1in 0in 0in 0in,
    clip]{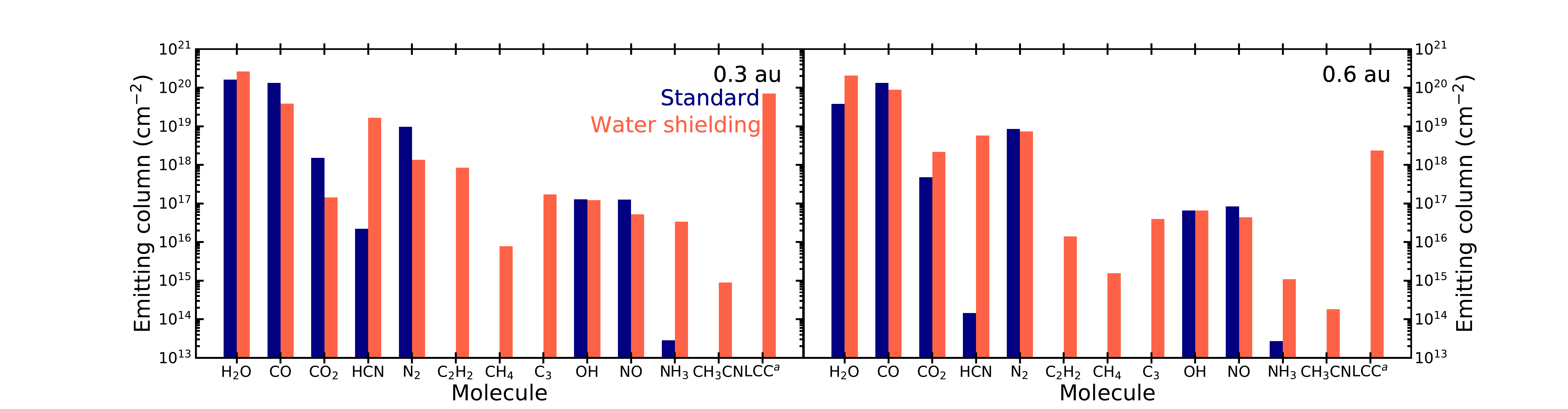}
    \caption{Abundant gas-phase species for the standard and water shielding models at 0.3 au and 0.6 au. \ce{C2H2}, \ce{CH4}, \ce{C3}, \ce{CH3CN}, and LCC have abundances below $10^{12}$ cm$^{-2}$ for the standard model. It should be noted that at 0.6 au, there is a higher abundance of LCC$^{a}$ in ice form. ($^{a}$ LCC signifies large carbon chains containing four or more carbons).}
    
    \label{fig:abundantspec}
\end{figure*}

    
\section{Discussion}
\subsection{Carbon}

    In Fig.~\ref{fig:mol_abuIRlayer}, the CO abundance is reduced within the \ce{CO2} iceline when water UV-shielding is included. The self-shielding of water lowers the abundance of water photo-products, such as atomic oxygen and OH, in the gas. These species are critical in the formation of CO, so its formation is slowed. This effect combined with dissociation reactions with He$^{+}$ in the upper atmosphere as discussed in Section~\ref{sec:Res} reduces the abundance of CO. Thus, more carbon is available for the formation of other species.
    
    Similarly to CO, we can also see that the abundance of \ce{CO2} is reduced when water self-shielding is included due to the atomic oxygen and OH-poor environment \citep[][]{Bosman2022CO2}; both are crucial to its formation. In contrast, we see increases in abundance for HCN, \ce{C2H2}, \ce{CH4}, and \ce{CH3CN}. 
    
    Figure~\ref{fig:abundantspec} shows that at 0.3 au, most of the carbon is incorporated into long carbon chains, such as \ce{C6H2} and \ce{C9H2} for the UV-shielding model with abundances of $3 \times 10^{-6}$ (relative to total H). This result is similar to \citet{Wei2019}, which explores releases of carbon from refractory carbon-rich grains. They find the efficient creation of large carbon chain species if carbon grain destruction is included. Though there are differences in the origin of the carbon chains between \citet{Wei2019} and our models, the similar end results suggests that inner disk chemistry drives large carbon chain production whenever there is little atomic oxygen available.

    Past models by \citet{Woitke2018} and \citet{Anderson2021} have found low abundances and fluxes of \ce{C2H2}. In their models, matching observation and theory requires a high C/O ratio, thus creating an inability to reproduce \ce{H2O} observations. Figure~\ref{fig:mol_abuIRlayer} shows that the \ce{C2H2} abundance is elevated in an {\em effectively} oxygen-poor environment, as created by UV-shielding in gas that is oxygen-rich (i.e. stellar O/H in content). As T$_{gas} >$ T$_{dust}$  in this region, this should lead to a strong increase in 13.7 $\mu$m \ce{C2H2} emission. At 0.3 au, our models produce a \ce{C2H2} column density of $8.4 \times 10^{17}$ cm$^{-2}$ in the region of the disk in the IR emitting layer. The models of \citet{Anderson2021} have found the column density of \ce{C2H2} to be in the range of $10^{14} - 10^{16}$ cm$^{-2}$, and \citet{Woitke2018} found it to be equal to $10^{17}$ cm$^{-2}$ for a C/O ratio of 0.46. It should be noted that most of this column is at a low gas temperature of 230K and thus only weakly contributes to any line emission. Our models thus produce a larger amount of \ce{C2H2} in higher regions of the disk without invoking a elevated C/O ratio.
    
    Comparing to the observations of \citet{Salyk2011}, they find best fit columns on the order of $10^{14} - 10^{15}$ cm$^{-2}$ for \ce{C2H2}. This is low compared to our value of $10^{18}$ cm$^{-2}$ at 0.3 au. This discrepancy could be explained by the way the column in derived in \citet{Salyk2011}, in which the emitting area for water was determined and applied it to all species. The region in which \ce{C2H2} has a high abundance is smaller than that of water, so it is expected that the \ce{C2H2} emitting region is smaller as well. This smaller emitting region would also imply a \ce{C2H2} excitation temperature that is higher than that of \ce{H2O} as is seen in \citet{Salyk2011}. If the actual emitting region of \ce{C2H2} is smaller than assumed in the fitting by \citet{Salyk2011}, then the column will have to be decreased significantly to compensate and still produce the same total flux. This could explain the mismatch between the observation derived columns and our predicted columns.

    Finally, it is important to note the carbon species that we expect to see in this region. We likely expect to observe \ce{CH4} with a column $8 \times 10^{15}$ cm$^{-2}$ at 0.3 au. High abundances of \ce{C3} and long carbon chains with low hydrogenation are seen in Fig.~\ref{fig:abundantspec} which indicates that these species could have detectable band emission. It is unlikely to detect \ce{C2H6} and \ce{C2H4} in this region as the models produced columns of less than $6 \times 10^{13}$ cm$^{-2}$ and $2 \times 10^{14}$ cm$^{-2}$, respectively. 

    \label{ssc:carbdisc}

\subsection{Nitrogen}
Water UV-shielding has a strong effect on the nitrogen-bearing species, increasing abundances of \ce{HCN}, \ce{NH3} and \ce{CH3CN} while lowering the abundance of NO. The abundance of \ce{HCN} is known to be sensitive to the gas phase C/O ratio \citep[e.g.][]{Cleeves2018}.  Our model effectively changes the C/O ratio and thus, HCN rises in abundance for the water UV-shielding model.  This is in large part driven by the chemistry discussed in Sec.~\ref{ssc:carbdisc}. However, the changes in \ce{NH3} and \ce{NO} imply that the active nitrogen chemistry is also changed. 

There seems to be three driving factors for the increased abundance of \ce{NH3}, \ce{HCN} and \ce{CH3CN}. The first driving factor is due to an impediment placed within the formation pathway of \ce{N2}. \ce{N2} is primarily formed from atomic N by the reactions:
\begin{equation}
\begin{split}
    \ce{N + OH -> NO + H} \\
    \ce{NO + N -> N2 + O}. 
\end{split}
\end{equation}
With the inclusion of water self-shielding the OH abundance is lowered and this channel is suppressed. Further, in both the standard and full models, the \ce{N2} formation through \ce{CN + N -> N2 + C} is suppressed by the competition with the \ce{CN + H2 -> HCN + H} reaction. The slow \ce{N2} formation leads to more nitrogen being available for species beyond \ce{N2}, most notably HCN and \ce{NH3}. This also impacts the abundance of \ce{NO} which is formed less efficiently in the UV-shielding model and thus has a lower abundance (Fig.~1). 

\ce{NH3} formation is initiated by the reaction of He$^{+}$ with \ce{N2} or HCN, forming N$^{+}$. The addition reaction with \ce{H2} allows for the eventual formation of \ce{NH4}$^{+}$, the precursor to \ce{NH3}. The main destruction channel for \ce{NH3} in these hot layers is atomic \ce{H}. Suppressing the photo-dissociation of \ce{H2O} lowers the production, and thus abundance of atomic \ce{H}. This increases the \ce{NH3} lifetime and thus abundance as more hydrogen is available in the form of \ce{H2}. We see that \ce{NH3} has a maximum abundance occurring at the water iceline with the inclusion of UV-shielding of roughly $10^{-7}$ , a value four orders of magnitude higher than when UV-shielding is excluded. 
 
Lastly, the active carbon chemistry allows for atomic nitrogen to react with the abundant carbon chains (\ce{C_xH}, \ce{C_yN}), which produces \ce{CN}, which reacts with \ce{H2} to form \ce{HCN}. \ce{HCN} can then react with the more abundant \ce{CH3+} to form \ce{CH3CNH+} the precursor for \ce{CH3CN}.  

The only nitrogen species that has so far been observed in the inner disk is HCN. HCN has been observed with columns of $10^{14} - 10^{15}$ cm$^{-2}$ in \citet{Salyk2011} and predicted to have columns of $10^{14} - 10^{16}$ cm$^{-2}$ in the line forming region at 0.3 au, by \citet{Woitke2018}. Though the column by \citet[][]{Woitke2018} matches observation, their emission line is weaker, similar to the case of \ce{C2H2} as the column is built up within deeper, cooler gas. Higher HCN fluxes are only seen by letting the C/O ratio approach unity in \citet[][]{Woitke2018}. 
 We produced a column of $10^{19}$ cm$^{-2}$ with the effects of UV-shielding included.  This column has a strong contribution from warm surface layers (e.g. Fig.~\ref{fig:2Dspecies} and thus the flux from our model is likely stronger than with the C/O = 0.46 from \citet{Woitke2018}. 

     The species that that might be detectable are NO and \ce{NH3} with columns of $5 \times 10^{16}$ cm$^{-2}$ and $3 \times 10^{16}$ cm$^{-2}$, respectively, while it is less likely to observe \ce{CH3CN} with a column of $9 \times 10^{14}$ cm$^{-2}$.

\label{sec:Dis}

\section{Conclusions}
In this work, we have studied the effects of water UV-shielding on the chemical compositions of the inner, planet forming region of protoplanetary disks. Specifically, we are looking at chemical abundances in the upper disk atmosphere where the IR line emission originates and the effects of water UV-shielding are most prominent. This is done in order to further our understanding of observed emission from organics which will be critical for the interpretation of observations by the James Webb Space Telescope.

We have concluded that water self-shielding has notable effects for hydrocarbons and nitriles. The lack of OH produced by \ce{H2O} dissociation suppresses \ce{N2}, \ce{CO}, \ce{CO2} formation. As a result, there is more carbon and nitrogen available for a rich carbon and nitrogen chemistry ($>$ 3 orders of magnitude), boosting the abundance of \ce{C2H2}, \ce{CH4}, HCN, \ce{CH3CN}, and \ce{NH3}.

The depletion seen in CO and \ce{CO2} cannot be explained alone by the formation of species such as HCN, \ce{C2H2}, \ce{CH4}, \ce{CH3CN}. A significant amount (up to 53 percent) of the total volatile carbon finds its way into larger carbon chains, such as \ce{C6H2} and \ce{C9H2}. The nitrogen released from \ce{N2} finds its way to HCN  with traces in \ce{CH3CN} and \ce{NH3}. We expect to observe \ce{CH4}, HCN, NO, and \ce{NH3} with column densities shown in Fig.~\ref{fig:abundantspec}.

The inclusion of water UV-shielding provides a way to increase the production of \ce{C2H2} and HCN which have been historically under-produced, without invoking a C/O ratio near unity. Model abundances for \ce{H2O} and \ce{CO2} are already in agreement with observation, noted in Paper \MakeUppercase{\romannumeral 1} and Paper \MakeUppercase{\romannumeral 3}, thus this is a  step forward in matching all four species at the same time.

Water UV-shielding is important for the entire chemical inventory. Through its ability to both block UV rays from penetrating deep into the disk and to create an effectively oxygen-poor environment, formation conditions become more favorable for various hydrocarbons and nitriles. This work has shown that water UV-shielding impacts the inner disk chemical composition and better reproduces observation. Thus, further studies that vary the C/O ratio with UV-shielding are needed to advance our understanding of the inner disk chemistry and its evolution.

\acknowledgments 

ADB and EAB acknowledge support from NSF Grant\#1907653 and NASA grant XRP 80NSSC20K0259. 
\software{Astropy \citep{astropy2013,astropy2018}, SciPy \citep{Virtanen2020},  NumPy \citep{van2011numpy}, Matplotlib \citep{Hunter2007}.}

\bibliographystyle{aa.bst}
\bibliography{Lit_list}

\appendix
\section{2D Structure Comparisons}
\label{app:appendix}
\begin{figure*}
\begin{minipage}[c]{\hsize}
    \includegraphics[width = \hsize, trim=0in 0in 0in 0in, clip]{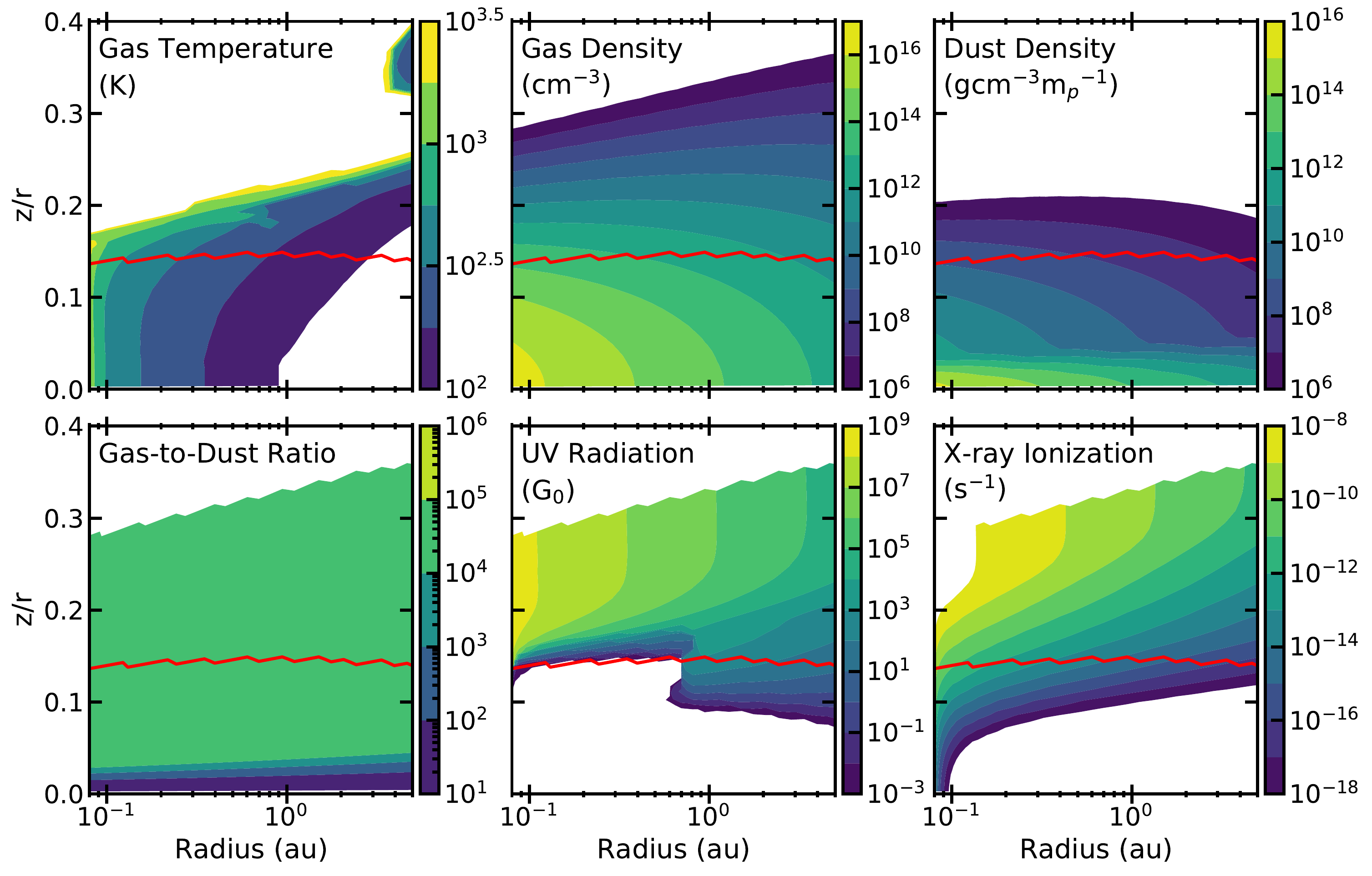}
\end{minipage}

    \caption{2D distributions of the inner 5 au of the disk showing the structure of the gas temperature and density, dust density, gas-to-dust ratio, and UV and X-ray radiation fields for the model with water UV-shielding. The UV radiation field is relative to the interstellar radiation field \citep[][]{Draine1978}. The red line at z/r $\sim$ 0.15 signifies our estimation of the location of the IR emitting layer. 
    }
    \label{fig:2D_9}
\end{figure*}

\begin{figure*}
\begin{minipage}[c]{\hsize}
    \includegraphics[width = \hsize, trim=1in 0in 1in 0in, clip]{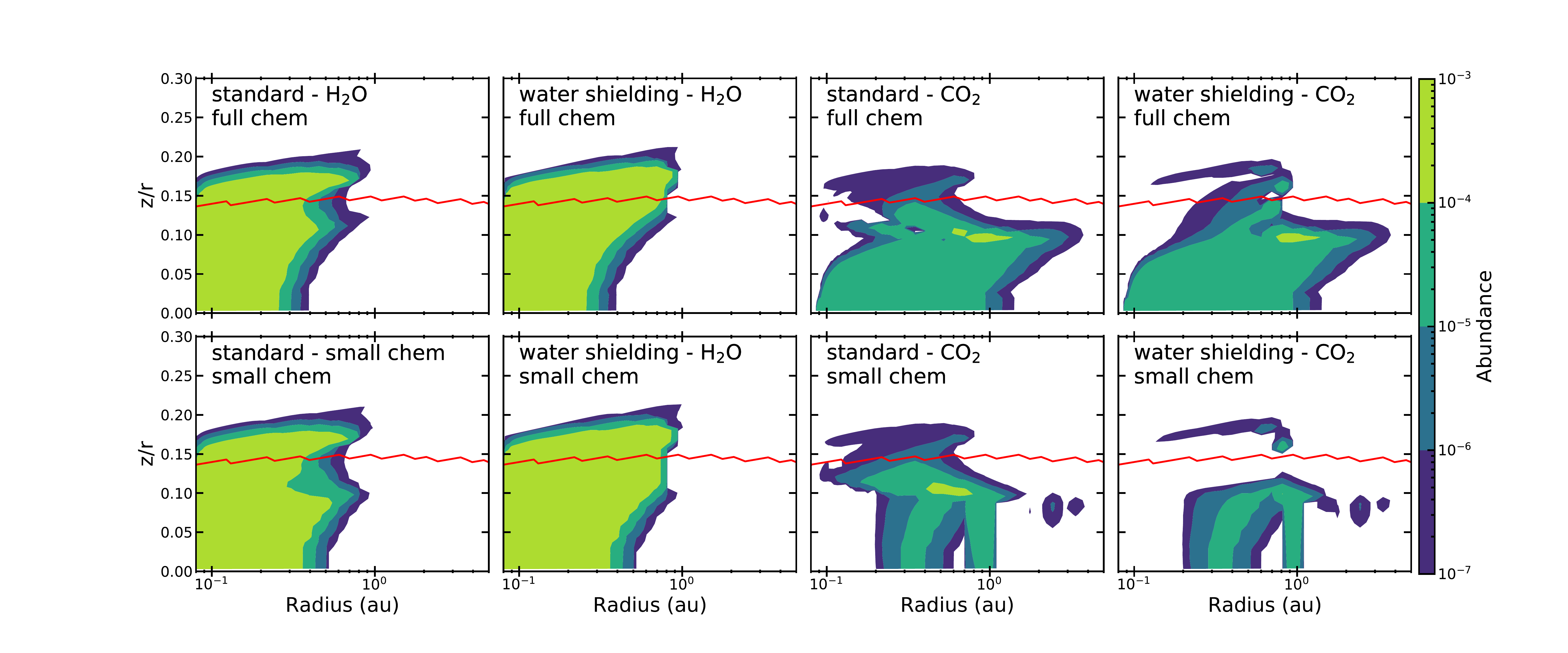}
\end{minipage}
    \caption{2D distributions of water and \ce{CO2} for a full and reduced chemical network shown in the top and bottom panels, respectively. The reduced chemical network is used in \citet{Bosman2022water}. The region above the red line at z/r $\sim$ 0.15 corresponds to the location of our estimation of the IR emitting layer. For water, we can see specifically around 1 au near the IR emitting layer that the model with UV-shielding has a noticeably larger abundance than the standard model, as expected. For a reduced chemical network, the abundances drop at closer in radii due to the fact that the water icelines are at smaller radii \citep{Bosman2022water}. We are focusing on the region in the IR emitting layer, so between the models, the abundance is about the same, validating the use of a smaller chemical network for obtaining the emitting area of water. For \ce{CO2}, we also see very similar abundances in the IR emitting layer between a full and reduced chemical network.}
    \label{fig:2D}
\end{figure*}

Figure~\ref{fig:2D_9} shows the 2D structure of the gas temperature and density, dust density, gas-to-dust ratio, and UV and X-ray radiation fields for the model with water UV-shielding. The UV radiation field is relative to the interstellar radiation field \citep[][]{Draine1978}. The red line at z/r $\sim$ 0.15 signifies the lower bound of the IR emitting layer. Figure~\ref{fig:2D} compares the 2D abundance structure of \ce{H2O} and \ce{CO2} for a full chemical network, as employed in this work, and a reduced chemical network, used in \citep{Bosman2022water}. We can see that for both the standard model and model with water UV-shielding, the resulting abundances in our estimation of the IR emitting layer for both species is independent of the chemical network used.

\end{document}